\documentclass[12pt,preprint]{aastex}
\usepackage{graphicx}

\shorttitle{EARLY LIGHT CURVE OF SN 2015F}
\shortauthors{Im, Choi, et al.}

\begin{document}


\title{The Very Early Light Curve of SN 2015F in NGC 2442:
A Possible Detection of Shock-Heated Cooling Emission and Constraints on SN Ia Progenitor System}


\author{Myungshin Im\altaffilmark{1,2,6}, Changsu Choi\altaffilmark{1,2,6}, 
Sung-Chul Yoon\altaffilmark{2}, Jae-Woo Kim\altaffilmark{1,2},  
 Shuhrat A. Ehgamberdiev\altaffilmark{3}, Libert A. G. Monard\altaffilmark{4}, 
\& Hyun-Il Sung\altaffilmark{5}}
\affil{\altaffilmark{1}Center for the Exploration of the Origin of the Universe (CEOU), Seoul National University, Seoul, Republic of Korea}
\affil{\altaffilmark{2} Astronomy Program, Department of Physics \& Astronomy, Seoul National University, Seoul, Republic of Korea}
\affil{\altaffilmark{3}Ulugh Beg Astronomical Institute, Tashkent, 
Uzbekistan}
\affil{\altaffilmark{4}
Kleinkaroo Observatory, Center for Backyard Astrophysics Kleinkaroo, Sint Helena 1B, PO Box 281, Calitzdorp 6660, South Africa}
\affil{\altaffilmark{5}Korea Astronomy and Space Science Institute, Daejeon 305-348, 
Republic of Korea}
\email{\altaffilmark{6}mim@astro.snu.ac.kr, changsu@astro.snu.ac.kr}




\begin{abstract}
 The main progenitor candidate of Type Ia supernovae (SN Ia)
 is white dwarfs in binary systems where the companion star is another 
 white dwarf (double degenerate system)    
 or a less evolved non-degenerate star with $R_{*} \gtrsim 0.1 ~R_{\odot}$ (single degenerate system),
  but no direct observational evidence exists that tells which progenitor system is more common.  
  Recent studies suggest that the light curve of a supernova shortly after its explosion can be used to set a limit on the progenitor size, $R_{*}$.    
 Here, we report a high cadence monitoring observation of SN 2015F, a normal SN Ia, in the galaxy NGC 2442 starting about 84 days before the first light time. With our daily cadence data, we catch the emergence of the radioactively powered light curve, but more importantly detect 
 with a $>$ 97.4\% confidence a possible dim precursor emission that appears at 
 roughly 1.5 days before the rise of the radioactively powered emission.    
  The signal is consistent with theoretical expectations for a progenitor system 
  involving a companion star with $R_{*}\simeq 0.1$ -- $1~R_{\odot}$ or  
  a prompt explosion of a double degenerate system, but inconsistent 
  with a typically invoked size of white dwarf progenitor of $R_{*} \sim 0.01 ~R_{\odot}$. 
   Upper limits on the precursor emission also constrain 
  the progenitor size to be $R_{*} \lesssim 0.1~ R_{\odot}$, 
  and a companion star size of $R_{*} \lesssim  1.0 ~R_{\odot}$, 
  excluding a very large companion star in the progenitor system.
  Additionally, we find that the distance to SN 2015F is $23.9 \pm 0.4$ Mpc. 
\end{abstract}


\section{INTRODUCTION}

 Explosive death of a star, supernova, is a dramatic cosmic event that 
 occurs at the end of the evolutionary path of many stars. 
 Observational study of such events can tell us whether our theoretical understanding
 of stellar evolution is correct, and thus studies of supernovae have been actively 
 pursued ever since modern astronomy has begun.

  Among various kinds of supernovae,  
 SNe Ia are believed to be the thermonuclear explosion of carbon-oxygen (C/O)  
 white dwarfs in close binary systems 
 (see Maoz et al. 2014 for a recent review on this subject). 
  This scenario is supported by the energetics and the chemical composition of
   the ejecta, as well as the progenitor ages inferred from their host
   galaxy properties 
  (Holye \& Fowler 1960; also see Hillebrandt \& Niemeyer 2000). 
 Yet we have no direct observational evidence that can constrain  
 their progenitor systems, such as the size of the exploding star
 and the nature of its companion. 
 Two progenitor systems are theorized. One is the Single Degenerate (SD) system where 
 the companion star is a non-degenerate star with a radius of 0.1 -- 100 $~R_{\odot}$ depending
 on its mass and evolutionary stage. The other is 
 the Double Degenerate (DD) system where two white dwarfs merge to produce a SN Ia
 in a close binary system as a result of orbital angular momentum loss via gravitational
 wave radiation.
  In both cases, the size of the exploding white dwarf is expected to be small 
 ($\lesssim 0.01 ~ R_{\odot}$). However, in the SD system, if the white dwarf was 
 a recurrent nova that acrretes hydrogen-rich material with a fairly high mass accretion
 rate ($dM/dt \sim 10^{-6}~M_{\odot}$/yr), the radius of the envelope could be as large as
 0.2 $R_{\odot}$ (e.g., Hachisu \& Kato 2003). In the DD scenario, the radius of the 
 shock breakout surface would be about 0.1 $R_{\odot}$ if a prompt detonation
 occurred as a result of the merger (Pakmor et al. 2012; Tanikawa et al. 2015), while
 in the case of a long-delayed ($> 10^{4}$ yr) explosion after the merger, the radius
 would be only about $0.01~R_{\odot}$ (Yoon et al. 2007).

  It has been suggested that the very early light curve within 
 $\sim1$ day of the explosion can constrain properties of the progenitor
 system such as the radius of the detonating star and the companion star
 (Nakar \& Sari 2010; Piro et al. 2010; Kasen 2010, hereafter, K10; Rabinak \& Waxman 2011, 
 hereafter RW11; Maeda et al. 2014; Kutsuna \& Shigeyama 2015). 
  This is because the luminosity from the 
 shock-heated materials that immediately follows the shock breakout is 
 proportional to the radius of the progenitor star for a given 
 set of progenitor mass and the supernova explosion energy 
 (K10; Piro et al. 2010; 
 RW11; Rabinak et al. 2012; Piro \& Nakar 2013, 2014; Pan et al. 2012).
  This phase has been shown to last for several
 hours or more, with the exact duration depending on the progenitor size
 (Rabinak et al. 2012; Piro \& Nakar 2013). The emission is generally dim with
 $M_{R} \sim -10$ -- $-14$ mag for a progenitor radius of $0.1$ -- $1 ~ R_{\odot}$ (Figure 3), 
 but it could be much brighter if a much larger star is involved ($M_{R} \sim - 17$ mag for
 $100 ~ R_{\odot}$; K10).

  Recently, several SNe Ia have been discovered in very early phases 
  (SN 2009ig: Foley et al. 2012; SN 2011fe: Nugent et al. 2011; 
  SN 2012cg: Silverman et al. 2012; SN 2012ht: Yamanaka et al. 2014; 
  SN 2013dy: Zheng et al. 2013; 
  SN 2014J: Zheng et al. 2014; Goobar et al. 2014; Olling et al. 2015; 
  Shappee et al. 2015),
 providing light curves within $\sim1$ -- 2 day of 
 the first light time\footnote{We use the term, ``first light time", 
 to indicate when the radioactively powered light curve started to rise, 
 and we distinguish the first light time   
 from the actual explosion time at around the shock breakout
 (e.g., Piro \& Nakar 2013).}
  and providing an opportunity to constrain the prognitor star radius.
 In general, studies involving 
     early light curves of normal SNe Ia suggest  
     small progenitor sizes.
       Using SN 2011fe, Nugent et al. (2011) constrained
 the progenitor radius, $R_{*}$, to be $R_{*} < 0.1 \,R_{\odot}$, while Bloom et al.
 (2012) suggested an even tighter constraint of $R_{*} < 0.02 \,R_{\odot}$
 based on a non-detection at 7.5 hours before the estimated first light time. 
  Similar attempts have been made using other SNe Ia with early light curves, giving 
 the limits of $R_{*} < 0.25$ -- $0.35 \, R_{\odot}$ for SN 2013dy (Zheng et al. 
 2013), $R_{*} < 1.5$ -- $ 2.7 \, R_{\odot}$ for SN 2012ht (Yamanaka et al. 2014),
 and $R_{*} < 0.34 \, R_{\odot}$ for ASASSN-14lp (Shappee et al. 2015). Similar limits are given for three SNe Ia from the {\it Kepler} mission (Olling et al. 2015). 
  On the other hand, Goobar et al. (2015) suggest that a progenitor system with 
  a scale of $R\gtrsim 1 ~R_{\odot}$ for SN 2014J. 
  For iPTF14atg, Cao et al. (2015)
  find an early UV light that is consistent with a system that has 
  a large companion star (several tens of $R_{\odot}$).  
     However, the SN studied in Cao et al. (2015) is
     an underluminous SN Ia, a peculiar one. 
   We also note a possible detection of a single progenitor system
   for SN 2012Z, another underluminous SN Ia (type Iax), in a pre-explosion
   image (McCully et al. 2014).

   These results assumed that the explosion time of SNe Ia coincides 
 with the first light time of the light curve 
 that is dominated by the heating due to radioactive decay of $^{56}$Ni.   
   However, the rise of the radioactively powered light curve 
 is expected to start a few hours to a few days after the explosion depending on how deep 
 the $^{56}$Ni layer is (Piro \& Nakar 2013), 
 while the UV/optical emission from shock-heated materials occurs almost immediately 
 after the explosion (e.g., H\"oflich \& Schaefer 2009).
  Hence, the previous constraints on SNe Ia progenitors have been put into question.
  This time gap between the thermonuclear
 explosion and the rise of the radioactively powered light curve has been
 called as ``dark phase" or ``delay time" which we note as 
 $t_{d}$ for convenience. The analysis of the early spectra of SN 2011fe by 
 Mazzali et al. (2014) led them to conclude that the actual explosion
 of SN 2011fe preceded about 1 day before the first light  
 time estimated by Nugent et al. (2011). The inclusion of the uncertainty 
 in the duration of the dark phase weakens the limits on the progenitor star 
 radius to $R_{*} \lesssim 0.1 R_{\odot}$ for SN 2011fe (Piro \& Nakar 2014; 
 Mazzali et al. 2014).
 
  The uncertainties in the estimate of the explosion time and $t_{d}$ 
  can be reduced by utilizing high cadence non-detection  
 data going back to several days prior to the first light time.
  For example, Mazzali et al. (2014) combined their estimate of the 
 explosion time and the non-detection of SN 2011fe around that time, 
 and put forward a constraint of 
  $R_{*} < 0.06 \, R_{\odot}$. This, however, depends on 
 the accuracy of the explosion time estimate again, and the constraint 
 on $R_{*}$ could be weaker if the SN 2011fe explosion occurred 
 even before their estimate. Unfortunately, there is a general lack of
 high cadence data prior to the first detection\footnote{Throughout this paper, 
  we use the term "first detection" to indicate the first detection of
  the radioactively powered light curve.} of SNe, so that 
 a study similar to this is not possible for the majority of SNe. 
   Ultimately, a direct detection of the emission from shock-heated 
  cooling prior to the first light time can lead to 
  a very tight constraint on both the explosion time and
  the progenitor property.

   SN 2015F was discovered by one of us (Monard) in NGC 2442 on
  2015 Mar.09.789 (UT) in its early phase, and has been 
  classified as a SN Ia (Fraser et al. 2015). The data from our regular monitoring program,
  Intensive Monitoring Survey of Nearby Galaxies (IMSNG),  
  of NGC 2442 show a pre-discovery detection at 2015 Mar.08.46. 
   We have been monitoring NGC 2442 on a daily basis from 2014 December 14,
   and the data reveal positive signals at Mar.05.57 and 
  Mar.6.55 that we interpret as an emission from shock-heated materials  
  as we show below. 
   In this paper, we will make use of these high cadence data to place  
  constraints on the SN Ia progenitor size.  In addition, we will derive  
  the distance modulus of NGC 2442 and the properties of SN 2015F.

\section{OBSERVATION}

  IMSNG is a high cadence imaging survey monitoring nearby galaxies to
 catch transients such as SNe in their early phases. The final goal of the survey
 cadence is 3 hrs, using a network of telescopes all over the world. We started our 
 test observations in 2013B with a 1m telescope at Mt. Lemmon Optical Astronomy 
 Observatory in US, SNUCAM on the 1.5m telescope at
 Maidanak observatory in Uzbekistan (Im et al. 2010), 
 and a 0.6m telescope of 
 Mt. Sobaek Optical Astronomy Observatory in Korea. 
 In 2014 October, we installed a 0.43m telescope (Lee 
 Sang Gak Telescope, LSGT hereafter; Im et al. 2015) at the Siding 
 Spring Observatory, Australia, which enabled us
 to perform a few hours to daily cadence observation of nearby galaxies 
 that are accessible in the southern hemisphere.
  Since 2014 December 14, $R$-band images of NGC 2442 have been taken once 
 to three times every night at a $\sim$1.5 hour interval 
 (weather permitting), using SBIG ST-10XME 
 camera on LSGT which provides a field of view of
 $17\farcm5 ~\times~11\farcm8$ and a pixel scale of $0\farcs48$. 
 Three frames, each with 180 s exposure, have been 
 taken at a given epoch. 
 Occasionally, $B$-band and $V$-band observations were added. 
  Typical seeing full-width-half-maximum (FWHM) values ranged 
  between $1\farcs8$ to $4\farcs0$ with a medium value at $2\farcs5$
  (Im et al. 2015).  
  The LSGT data were reduced with a standard procedure of dark subtraction and
  flat-fielding as soon as the data were taken. 
  
\begin{figure}[p]
\center{
\includegraphics[width=14cm]{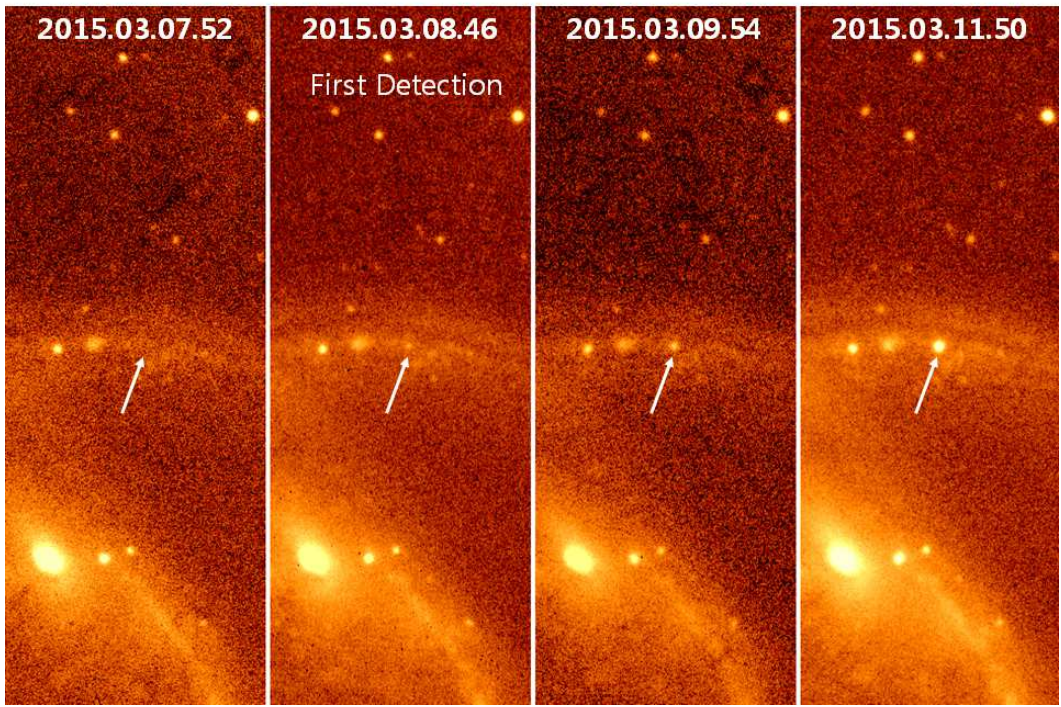}}
\caption{The $R$-band images of SN 2015F taken with LSGT, before the first 
detection of the radioactively powered emission (the leftmost) and after. 
Indicated in the figure are the UT-date of the observation, and the location of SN 2015F.}\label{fig1}
\end{figure}

   During the monitoring observation, the emergence of SN 2015F was recorded
  in the LSGT images taken at 2015 Mar. 08.46, 09.54, and 11.50 (Figure \ref{fig1}). 
   In comparison, we show an image taken at 2015 Mar. 07.52, approximately 
  one day before the first detection of
  SN 2015F. To better identify the SN by subtracting a reference image, 
  we constructed a master reference image using the LSGT data taken during
  2015 January to February under good weather conditions.
   The reference image was convolved with a Gaussian profile 
   and flux-scaled to match the seeing and the zero-point of
   each epoch image. And then, the reference image
   was subtracted to yield a difference image.  
   Some LSGT data had better seeing than the reference image. 
   In such cases, we convolved the later epoch data to match the reference image 
   resolution. The subtraction of the reference image removes NGC 2442 effectively, 
  allowing us to see SN 2015F clearly.  

    The photometry calibration was done using calibration stars from the data release 8
   (DR8) of the AAVSO photometric all-sky survey (APASS; Henden et al. 2012). 
   The $B$ and $V$-band data were taken directly from
  APASS values, and the $R$-band values of calibration stars were obtained
  by converting APASS $r$ and $i$-band photometry to the Johnson $R$-system
  using a SDSS photometry transformation equation in the form of
  $R = r - 0.2936 \times (r - i) - 0.1439$ with a dispersion of $\sigma=0.007$ mag. 
   This procedure gives photometry zero-point for each epoch data, with a typical
  zero-point error of 0.02 mag.
   The photometry was done by running SExtractor (Bertin \& Arnouts 1996) on images before 
   (for calibration stars) and after (for SN 2015F) the subtraction of the reference 
   image. We used $3\farcs0$ diameter aperture and applied an aperture correction 
 that was derived from  stars in the vicinity. The use of this size of aperture 
 has an advantage  of minimizing residual fluxes of extended features 
 coming from an imperfect galaxy
  subtraction, if present. Upper limits at 3$-\sigma$ are adopted for 
  non-detections.

   The discovery image and subsequent images of SN 2015F were also taken
  with a SBIG-ST8-XME camera (no filter) on Meade 12 or 14 inch RCX400 telescopes 
  at the Kleinkaroo observatory in South Africa.   
  Stacked images of three to four 13 s frames 
  were used to search for SNe, and more than ten 13 s images were 
  taken during the follow-up observation each night. 
    The images were first calibrated with a standard reduction procedure of dark subtraction
  and flat-fielding and stacked to create a deeper image. An image of NGC 2442 taken before
  the SN detection was subtracted to create a difference image on which the 
  photometry was performed in the same way as the LSGT data.  
   For these data, color transformation equations were derived 
  using $B-V$ colors of stars in the NGC 2442 field to convert
  the clear magnitudes to $R$-band magnitudes. 
  The transformation equation has the form of 
  $R = C_{1} + C_{2} \times (B-V)$. The $C_{1}$ and $C_{2}$ values were
  determined for each image and typical $C_{2}$ values are 
  $-0.1$ and $-0.23$ for the 12 inch and 14 inch telescopes, respectively.
   For the $B-V$ color of the supernova, we first took the dereddened $B-V$ values 
  of SN 2011fe at the same number of days from the $B-$band maximum 
  (Richmond \& Smith 2012), and then reddened them using 
  the Galactic and internal (host galaxy) extinctions toward SN 2015F where the internal extinction
  comes from an analysis of the long term light curve fit (Section 3.1). 
   The resulting $B-V$ values range from 0.58 (Mar.09) to 0.23 (Mar.15) mag.
   At the $B$-band maximum, the $B-V$ value estimated this way is 0.2 mag, which 
  is in excellent agreement with the observed $B-V$ color of 0.2 mag for SN 2015F.  
    Taking into account of the fitting error for the transformation equation 
   and the uncertainty in $B-V$ (taken to be 0.1 mag), we get an uncertainty in  
  the transformed $R$-band magnitude of 0.05 -- 0.1 mag. 
   Note that $B-V$ color could be redder by 0.3 mag in the early phase,
  since SN 2442 is not exactly identical to SN 2011fe (e.g., Yamanaka et al. 2014), 
  but such an offset changes the resultant
  $R$-band magnitude only by 0.03 -- 0.07 mag.

\section{LIGHT CURVE ANALYSIS}

\subsection{Long-term Light Curve and Distance to SN 2015F}

   Figure 2 shows the long-term $BVR$ light curve of SN 2015F up to 22 days 
  after the $B$-band maximum. Also plotted are the light curves  
  of SN 2011fe
  (Munari et al. 2013; Vinko et al. 2012; Richmond \& Smith 2012). Here,
  the SN2011fe light curves are shifted in y-axis so that they correspond to  
  the distance and the Galactic and internal extinctions of SN 2015F (Table 1).  
   For SN 2011fe, we assumed the distance modulus of 
  $\mu = 29.28$\footnote{Estimates of $\mu$  
  vary from 29.04 to 29.53 (Feldmeir et al. 1996; Macri et al. 2001; Sakai et al. 2004; Rizzi et al. 2007; Shappee \& Stanek 2011; Richmond \& Smith 2012; Vinko et al. 2012; Munari et al. 2013; Lee \& Jang 2012; Tammann \& Reindl 2013), and our adopted value of 
  29.28 corresponds roughly to a mid-point of
  these estimates.} (Lee \& Jang 2012), the epoch of maximum brightness in $B$-band of 
  $t_{\rm{max}}(B) = 2,455,815.00$ JD (average of $t_{\rm{max}}$ from 
  Vinko et al. 2012, Munari et al. 2013, Tsvetkov et al. 2013, and
  Pereira et al. 2013), the Galactic reddening of 
  $E(B-V) = 0.008$ (Schlafly \& Finkbeiner 2011), and 
  the host reddening of $E(B-V)_{\rm{host}} = 0.025$ (Patat et al. 2013).      
 After adjustment, the light curves of SN 2011fe and SN 2015F match closely.
 This suggests that SN 2015F is a SN Ia very similar to SN 2011fe.     
  
   In order to characterize SN 2015F in more detail, 
   we fitted the $BVR$ light curve from LSGT with  
   the MLCS2k2 model (Jha et al. 2007) using the SNANA software v10 (Kessler et al. 2009). 
    The Galactic reddening of $E(B-V) = 0.179$ mag is adopted from
   Schlafly \& Finkbeiner (2011; $A_{R}=0.440$ mag, $A_{B} = 0.735$ mag, and
  $A_{V} = 0.556$ mag). The fit returns quantities such as 
  the distance modulus and the host galaxy extinction ($A_{V}$), 
  and other parameters such as $\Delta$ (the stretch factor in the MLCS2k2 model)
  and $t_{\rm{max}}(B)$.

The first two of these quantities, distance modulus and host galaxy extinction,
play an important role in our analysis of the early light curves and constraining the progenitor size.    
     Importantly, prior to this work,  
     the distance to NGC 2442 has been 
   poorly known, with published distance 
   values varying from 17 Mpc (Tully 1988) to a more updated value of the distance modulus  
   $\mu =$ 31.66$\pm0.17$ ($21.5 \pm 1.7$ Mpc) that is based on the group distance (Tully et al. 2009).

  The fitting results are summarized in Table 1. The values and errors of the quantities
  such as $t_{\rm{max}}(B)$,
  $\Delta$, $B_{\rm{max}}$ ($B$ magnitude at maximum brightness), and $\mu$, 
  are direct outputs from the SNANA software.
   Additionally, we derived other parameters by combining the SNANA output values with
 external information. 
   The derived parameters are $\Delta m_{15}$ ($B$ magnitude difference between the maximum
 brightness and the brightness at 15 days after the maximum), $M_{B,\rm{max}}$ (absolute
 $B$ magnitude at maximum brightness), $t_{\rm{rise}}$ (days between the first light time and
 $t_{\rm{max}}(B)$), and $E(B-V)_{\rm{host}}$. Errors of these quantities
 are taken as the square root of the quadratic sum of errors of the quantities 
 involved in the derivation. 
  Note that when evaluating $t_{\rm{rise}}$, we assumed a mid-point of first light times
 derived from two fitting methods (Section 3.2).

   Our fitting gives the distance modulus of $\mu = 31.89\pm 0.04$,
  the host galaxy reddening parameter of $E(B-V) = 0.035 \pm 0.033$ mag, 
  $\Delta m_{15} = 1.26 \pm 0.10$, and $t_{\rm{max}} = 2,457,106.48 \pm 0.09$ JD 
   (2015 Mar.24.98, UT).  We conclude that the distance to NGC 2442 is $23.88 \pm 0.40$ Mpc, 
  somewhat larger than previous estimates,  
  and the dust-extinction by the host galaxy is small. 
   The fitting results also confirm
  that SN 2015F is very similar to SN 2011fe 
  which has $\Delta m_{15} \simeq 1.1$ and $M_{B,\rm{max}} \sim -19.4$, 
  showing that SN 2015F is a typical SN Ia.

\subsection{Early Light Curve and First Light Epoch}

   Figure 3 shows the early light curve (-5 to +10 days after the first detection) 
   of SN 2015F. 
   Plotted along with the data points are the results from a single power-law fit 
   (thick solid line),
  and a broken power-law fit (thin solid line), and the model predictions for the precursor
  emission from shock-heated materials (see below for the fitting functions and
  Section 4 for explanations for the models).
   Table 2 shows the photometry result of the early light curve from
    -84 days to 8 days.
    The values presented in Table 2 are not corrected for the 
   Galactic extinction, which is $A_{R} = 0.44$ mag.
   The AB magnitude offset of 0.22 mag is used for the model prediction 
   (e.g., Jeon et al. 2010).

   The early light curve up to 8 days after the first detection was fitted with two functions. 
  One is a single power-law ($(t-t_{0})^{\alpha}$) where $t_{0}$ is the first light time  with 
  respect to the first detection epoch, and $\alpha$ is the power-law index. Another is   
  a broken power-law function as given below (e.g., Zheng et al. 2013): 
\begin{equation}
 F(t) = F_{0} \,[(\frac{t-t_{0}}{t_{b}})^{\alpha_{1}} + (\frac{t-t_{0}}{t_{b}})^{s(\alpha_{1} - \alpha_{2})}]^{-1/s}, 
\end{equation}
where $F$ is the flux, $F_{0}$ is the normalization constant of the flux,
$t_{b}$ is the break time, $\alpha_{1}$ and
$\alpha_{2}$ are the power-law indexes before and after the break, and $s$ is a  smoothing
parameter. When $s=-1$, Eq. (1) reduces to a simpler broken power-law function that has 
been often used to model gamma ray burst afterglow (e.g., Urata et al. 2009).
 Zheng et al. (2013, 2014) show that a broken power-law function 
 can fit the very early-light curve of SN 2013dy and SN 2014J, when
 there is a data point at $\sim$ 0.5 days within the first light time.

  The single power-law fit gives the result of $t_{0} = 1.61 \pm 0.10$ days
 and $\alpha = 2.32 \pm 0.05$, with a reduced $\chi^{2}$ value of 
 $\chi^{2}_{\nu} = 0.79$. When using a fixed value of
 $\alpha = 2$,  we get $t_{0} = 1.01 \pm 0.02$ days, but with $\chi^{2}_{\nu} = 3.72$.
 
  On the other hand, we find that the best-fit result of the broken power-law fit converges
 to the single power-law fitting result. This is because we cannot constrain 
 $\alpha_{1}$ and $s$ values effectively due to the lack of a deep data point at 
 $\lesssim 0.5$ days before our first detection. Even so, 
  we can try to fit the existing data by fixing $\alpha_{1}$ and $s$ to the values  
 obtained for another SN. Adopting the best-fit values of $\alpha_{1} = 0.88$ and $s=-6.32$ of 
 SN 2013dy (Zheng et al. 2013), we obtain $t_{0} = 0.39 \pm 0.04$ days, 
 $t_{b} = 1.98 \pm 0.06$ days, 
 and $\alpha_{2} = 1.89 \pm 0.02$ with $\chi^{2}_{\nu} = 0.97$, which is a reasonably 
 good fit. The broken power-law fitting result in Figure 3 uses these parameters.
   One can  obtain $t_{0} \sim 0.1$ day at $\chi^{2}_{\nu} \sim 1.2$ 
 by adopting $\alpha_{1} \simeq 0.5$ and $s \simeq -250$, 
 although such a case produces a bumpy feature in the light curve near the first detection
 that looks artificial.
 These results suggest that the rise time is about 16.9 to 18.1 days.
  
  We adopt the result of a single power-law fit for the following discussion,
 since it gives the best fit to our data. However, we should bear in mind that 
 the first light time can be much smaller than the value inferred from the single
 power-law fit as was the case for SN 2013dy (Zheng et al. 2013) and 
 SN 2014J (Goobar et al. 2014).
 
 In the previous section, we mentioned that the uncertainty in $B-V$ color in 
the early light curve may cause a systematic offset of 0.03 -- 0.07 mag 
through the transformation of clear magnitudes to $R$-band magnitude 
for the South African
data. We find that the small offset in the photometry does not alter the fitting results much (well within the error estimates). This is because that the LSGT data
are the most constraining data for the light curve fitting, and that the small systematic offset does not exceed errors of the South African photometry 
data points.

\section{POSSIBLE DETECTION OF EMISSION FROM SHOCK-HEATED MATERIALS}

   To search for possible precursor UV/optical emission from shock-heated materials,
   we analyzed all the imaging data before the first detection that go back to 
   2014 December 14. The images taken at each night were stacked together to create a 
  deeper image. Through this process, we created stacked images at 40 epochs before the
  first detection.
   An aperture photometry with a $3\farcs0$ diameter was performed at the position of SN 2015F 
   in the difference images at each epoch. 
  
   The result is presented in Figure 4. Interestingly, we find that the measurements from two epochs, at 
  3 and 2 days before the first detection, show weak but positive signals 
  at 2$-\sigma$ significance (or $R \sim 21$ AB mag). 
  The significance of the combined signal from the two epochs is 3$-\sigma$, 
  and thus the formal probability of this being a true detection is 99.7\%.

    However, we caution that any unrecognized instrumental effects could 
    produce a spurious signal 
   at this level. Therefore, we varied our image reduction and analysis methods 
  by choosing different sets of images to construct a reference image 
  and performing additional flat-fielding using skyflats made from images of other
  targets observed during the same period. We also adopted various background annuli 
  for the photometry. Even after these changes, the signal persisted 
  at a similar statistical significance. 
   We checked the detector temperatures during the monitoring observation,
  and  we find them to be stable over the course of the monitoring period, 
  i.e., no anomaly in the detector temperature during these two nights. 
  SN 2015F was placed at different locations on the chip in each time it was
  observed, but no hot/warm pixels or other detector defects were found 
  in these positions including the Mar.05 and Mar.06 data.

   Spurious, non-astrophysical signals can potentially be identified
     by examining source FWHM values. The FWHM values of 
     astrophysical sources should be similar to those of stars in the same
     image. As a test, we inserted artificial stars at random locations in the
     images 
     with the same flux as our potential detection and the seeing FWHM
     values that match the stellar FWHM values 
     ($4\farcs0$ and $2\farcs6$ for Mar.05 and Mar.06 
     respectively). We find no significant difference between the injections 
     and our detection in size or shape, although we caution that 
     the test is not definitive in this case due to the low detection
     S/N.

   Given that no 2$-\sigma$ signal was detected over two consecutive nights 
  in the other 38 consecutive epoch-pairs, we suggest that this is not a spurious
  signal. Based on the 39 consecutive epoch-pair measurements, we set a conservative 
  probability of 2.6\% (1/39) for two consecutive 2$-\sigma$ detections 
  occurring randomly at a particular time window. Therefore, we consider
  that the probability of this signal being real is $>$ 97.4\%.

      If the signal is due to the shock-heated cooling emission, 
   the dark phase period corresponds
    to $\sim 1.5$ days.  If the signal at $t\simeq -3$ days is due to a random
   coincidence, then the dark phase period would be $\sim 0.5$ days.
   Dark phase periods have been noted for SN 2010jn 
   ($t_{d}$=1.0 day, Hachinger et al. 2013), SN 2009ig ($t_{d} \simeq 1.6$ day; Piro \& 
   Nakar 2014) and SN 2011fe 
   ($t_{d}$ = 0.5 to 1.5 day; Mazzali et al. 2014; Piro \& Nakar 2014), 
   with respect to the first light time estimates from a single power-law 
   or a $t^{2}$ light curve. Therefore, the inferred dark period of
   $t_{d} \sim 1.5$ days is reasonable, given the dark phase estimates for the other SNe Ia
   and theoretical expectations. 

    To constrain the progenitor property with this possible shock-heated cooling signal,
   we use analytic models of the bolometric light
   curve and the temperature curve by RW11 and K10. The model of 
   RW11 describes the evolution of the early shock-heated cooling 
   emission from the progenitor itself, while the model of K10 is for the 
   emission arising from an interaction between the companion star and the ejected materials.
    To model the $R$-band light curve, we used a black body radiation spectrum of 
   a given effective temperature, and used the flux at the effective wavelength
   of $R$-band (0.65 $\mu$m). 
   
 For the RW11 model,   
\begin{eqnarray}
 L(t) & = & 1.2 \times 10^{40} ~\frac{R_{10} E^{0.85}_{51}}
  {M_{c}^{0.69} \kappa^{0.85}_{0.2} f_{\rm{p}}^{0.16}} ~ t^{-0.31}_{d} ~ \rm{erg\,s^{-1}} \nonumber
\\
 T_{eff}(t) & = &4.1 \times 10^{3} ~\frac{R_{10}^{1/4} E^{0.016}_{51} M^{0.03}_{c} \kappa_{0.2}^{0.27}}
  {f_{\rm{p}}^{0.022}} ~ t_{day}^{-0.47} ~ \rm{K}.
\end{eqnarray}

 For the K10 model,
\begin{eqnarray}
 L(t) & = & 2.0 \times 10^{40} ~\frac{R_{10} M_{c}^{1/4} v_{9}^{7/4}}{\kappa_{0.2}^{3/4}} ~
   t_{day}^{-0.5} ~ \rm{erg\,s^{-1}}\nonumber
\\
 T_{eff}(t) & = & 5.3 \times 10^{3} ~\frac{R_{10}^{1/4}}{\kappa_{0.2}^{35/36}} ~t_{day}^{-37/72}
   ~ \rm{K}.      
\end{eqnarray}
 
    Here, $R_{10}$ is the radius of the progenitor or the companion star
    ($R_{*}$ = (separation distance)/2 for the K10 model) in units of $10^{10}$ cm, 
   $E_{51}$ is the explosion energy $E=E/10^{51}\,\rm{erg}$,
   $M_{c}$ is the progenitor (or the ejecta) mass in units of 1.4 $M_{\odot}$, 
    $\kappa_{0.2}$ is the opacity in units of $0.2\,\rm{cm^{2}}\,\rm{g^{-1}}$, 
    $f_{\rm{p}}$ is the form factor that ranges between 0.031 and 0.13 (Calzabara 
    \& Matzner 2004; RW11), and $v_{9}$ is the expansion 
    velocity of the ejecta in units of $10^{9} \rm{cm\,s^{-1}}$. For the RW11 model,
    we adopt $M_{c} = 1/1.4$, $E_{51}=M_{c}$, $\kappa_{0.2} = 1$, and 
    $f_{\rm{p}}=0.05$. For the K10 model, we adopt $M_{c}=1/1.4$, $\kappa_{0.2}=1$, 
    and $v_{9}=1$.
               Note that the K10 prediction is anisotropic and the strength of 
     the signal varies with viewing angle.
     The K10 curve in Eq. (3) describes the case with a viewing angle that 
    gives nearly maximal observable signal of the shock-heated cooling emission
    (``optimal viewing angle"). We define the "common viewing angle" as 
    the angle that corresponds roughly to an 80 percentile of the viewing angle 
    where the observed signal is 10\% of the signal under the optimal viewing
    angle (Bloom et al. 2012). To model the common viewing angle case, we simply
    scaled the optimal viewing angle prediction by a factor of 10.    
    
     These theoretical light curves are plotted in Figure 3 for several 
   different values of $t_{d}$ that range from 0.5 to 2.5 days, and for 
   a range of $R_{*}$ values. In Figure 4, the theoretical light curves are
   plotted assuming $t_{d}=1.6$ days, where the RW11 model is for
   0.1 $R_{\odot}$ (the dashed lines), and the K10 model (the dotted lines) is for 
   $0.1 R_{\odot}$ (an optimal viewing angle), 
   $1 R_{\odot}$ and $2 R_{\odot}$ (a common viewing angle). 

    We find that the observed signal can be explained with a variety of progenitor systems.
    With the RW11 model, a progenitor radius of $\sim 0.1 R_{\odot}$ fits the data well, 
   which is consistent with a DD system that undergoes a prompt detonation 
   (Pakmor et al. 2012; Tanikawa et al. 2015) and a SD system where the 
   white dwarf is a recurrent nova with rapid mass accretion of hydrogen-rich matter
   (e.g., Hachisu \& Kato 2003). 
   On the other hand, this result is inconsistent
   with a DD system with a long-delayed explosion (Yoon et al. 2007) and with
   a SD system where the white dwarf radius is predicted as small as 
    $R_{*} \lesssim 0.01 R_{\odot}$ at the shock breakout (Hoflich et al. 2009; Yoon \& Langer 2004; Woosley \& Kasen 2011).  
    For the K10 model, the result can be explained by a SD system with a companion star 
   with $R_{*}=0.1$ -- $1 ~R_{\odot}$, where $R_{*}=0.1~R_{\odot}$ is for an optimal viewing
   angle, and $R_{*} \sim 1~R_{\odot}$ for a more common viewing angle. This result excludes
   cases where a companion star has $R_{*} \gg 1~R_{\odot}$ like that of 
   a subgiant of a red giant star.  
   
\section{UPPER LIMITS ON $R_{*}$}
         
     Clearly, our constraints on $R_{*}$ depends critically on how significant the
    detected signal is. Alternatively, we can take    
    a more conservative approach of considering 3$-\sigma$ upper limits only to 
    place a limit on $R_{*}$. 
    With the upper limits that extend to 
    several days before the first detection, we can still provide 
    a meaningful constraint on the progenitor system that is largely independent of
    $t_{d}$.

    Using Eqs. (2) and (3), we calculated $R_{*}$ values for 3$-\sigma$ limits
   at epochs that follows an assumed explosion time. The smallest value 
   among such $R_{*}$ values, is taken to be an upper limit on $R_{*}$, $R_{*,\rm{up}}$.
    When $R_{*} \lesssim 1 ~R_{\odot}$, the light curve peak of the shock-heated cooling emission appears 
   within 1 day of the explosion (Figure 4), so $R_{*,\rm{up}}$ is determined by a 3$-\sigma$
   upper limit that is closest in time to the assumed explosion time in such a case.

     We also need to consider $R_{*,{\rm min}}$, the minimal 
    radius that one can probe with a given set of time series observation. 
     This is necessary since the shock-heated cooling emission can drop suddenly 
    when the diffusion wave reaches material where the gas pressure dominates
    the radiation pressure (Rabinak et al. 2011, 2012). This limits the duration of the shock-heated
    cooling curve in such a way that the larger the progenitor is, the longer the duration becomes. 
     For a given time $t$ since the explosion, the size of the progenitor for which 
    the light curve drop suddenly at $t$ can be given as  
    \begin{equation}
    R_{*,\rm{min}} \simeq 0.013 \,E_{51}^{-0.66} \,M_{c}^{0.56}\, f_{0.05}^{0.15}\, t_{4hr} ~R_{\odot},
    \end{equation}
    where $t_{4hr}$ is time in units of 4hr (RW11; Bloom et al. 2012). 
     For one day cadence observation, we are limited to probing 
   $R_{*} > R_{*,\rm{min}} \sim 0.07\, R_{\odot}$.

       In Figure 5, we show $R_{*,{\rm up}}$,  
    and $R_{*,\rm{min}}$ 
    as a function of the assumed explosion time for which 
    the time of the first detection of the radioactively powered
    light curve is set to 0. For $R_{*,\rm{up}}$,
    we used the RW11 model (the dashed line), and the K10 model for two cases -- one for an optimal
    viewing angle to detect the shock-heated cooling emission (the dotted line),
    and another for a much less optimal viewing angle that produces about 
    10\% of the observed emission under an optimal viewing angle (the dot-dashed line).     
     At a given time, the larger value between
   $R_{*,\rm{up}}$ and $R_{*,\rm{min}}$ can be taken as the constraint 
   on the progenitor size.    
     The figure shows sawtooth-like curves for both $R_{*,\rm{max}}$ and $R_{*,\rm{min}}$,
   which result from the cadence of our data points. The constraint on $R_{*}$ is
    the strongest when the explosion time is near our data points in such a way that 
    the maximum of the light curve coincides with the epochs where the data were taken. 
      As we noted earlier, the explosion time is most likely at between the first light
    time of a single power-law fit and a few days before it. 
      In such cases, we find that $R_{*}$ stays at 
   $R_{*} \lesssim 0.1 \, R_{\odot}$ for fluxes coming from  
   a progenitor star (RW11) or interaction with a companion star at an optimal
   viewing angle. For the K10 model with a less optimal viewing angle, we find that 
   $R_{*} \lesssim 1 \, R_{\odot}$. 
    If the explosion time falls between the first detection time and
  one day before it, then the constraint on the progenitor system becomes about 10 times
  weaker. These limits are in agreement with the inferred progenitor size from the possible
  shock-heated cooling emission as discussed in the previous section.

    The constraint of $R_{*} \lesssim 0.1 \, R_{\odot}$ excludes many possible SN Ia 
    progenitors. As discussed in Bloom et al. (2012), 
    H-burning or He-burning MS stars with $M = 0.5 - 3.0 M_{\odot}$ 
    would have radii of $R_{*} > 0.2\,  R_{\odot}$ and can be
  excluded. C-burning MS stars are expected to have $M > M_{\odot}$ (Boozer et al. 1973), 
  and many such stars can be excluded when $M > 2 \, M_{\odot}$.

    A useful constraint on a companion star can be obtained too. Our result suggests 
   that the radius of a companion star should be $R_{*} \lesssim R_{\odot}$
   even if the viewing angle is not in an optimal direction. 
    This excludes red giant stars with 1-2 $M_{\odot}$ that would have
  radii $R_{*} \gtrsim 100\,R_{\odot}$. MS sub-giants with 5 -- 6 $M_{\odot}$  
  would have radii $R_{*} \sim 10 \, R_{\odot}$, and they can be excluded too. 
  MS stars with $\sim 1 \, M_{\odot}$ are consistent with our limit.

\section{CONCLUSION} 

  In this paper, we presented the light curve of a SN Ia, 
  SN 2015F, between -84 and 41 days with respect to the first light time. 
   Our data caught the rise of SN 2015F at a daily cadence, 
  providing an estimate of the first light time of 
  1.6 days before the first detection (single power-law fit). 
    Through our light curve analysis, we determined the distance to NGC 2442 
 and SN 2015F to be $23.9 \pm 0.4$ Mpc, and 
 the reddening parameter of $E(B-V) = 0.035 \pm 0.033$. 

   More importantly, we detected a possible signal from shock-heated cooling emission at 
  -3 and -2 days prior to the first detection of the radioactively powered light. 
  Additionally, we obtained upper limits for any precursor 
  emission over 40 nights before the first detection. 
   The possible detection of the shock-heated cooling  
  emission places stringent limits on the SN 2015F progenitor, 
  allowing a $R_{*} \sim 0.1 \, R_{\odot}$ progenitor system (such as in a prompt detonation
  of a DD system), or companion stars with $R_{*} \simeq 0.1$ -- $1\,R_{\odot}$ in a SD system.
   On the other hand, the possible detection and the upper limits around 
  and before the first light time exclude a very small progenitor with $R_{*} \ll 0.1 \, R_{\odot}$, 
  large progenitors with $R_{*} \gg 0.1 R_{\odot}$ or 
  a very large companion star with $R_{*} \gg R_{\odot}$, 
  and these conclusions are largely independent on the exact time of the explosion time.

   The detected shock-heated cooling emission is at a level of 
   $R=21$ mag at $\sim20$ Mpc and with various kinds of extinctions along the line of sight ($M_{R}= -11$ mag).
    This detection, although marginal ($>$ 97.4\%), 
    indicates that high cadence observation 
  of SNe Ia is a very promising way to probe their progenitor systems that have been
  elusive in the previous searches.    
     A secure detection of such a signal
   requires a sensitivity only a few times better than this work, 
   which can be easily achieved with observations using 1-m class telescopes (e.g., Lee et al. 2010).

\acknowledgements
This work was supported by the Creative Initiative program, No. 2008-0060544, of the National Research Foundation of Korea (NRFK) funded by the Korean government (MSIP). 
This paper includes the data taken at the Siding 
Spring Observatory in Australia. We thank the anonymous referee for useful comments.

{}

\clearpage

\begin{deluxetable}{cccccccc}
\tablecaption{Long-term Light Curve Fitting Results}
\tablewidth{0pt}
\rotate
\tablehead{
\colhead{$t_{\rm{max}}(B)$} & \colhead{$\Delta$} & \colhead{$\Delta m_{15}(B)$} &  
\colhead{$B_{\rm{max}}$} & \colhead{$M_{B,\rm{max}}$} & \colhead{$\mu$} &
\colhead{$t_{rise}$} & \colhead{$E(B-V)_{\rm{host}}$}  \\
\colhead{(JD)}      & \colhead{}    &  \colhead{(Mag)}  &
\colhead{(Mag)}      &  \colhead{(Mag)}   &  \colhead{}  &
\colhead{(Days)}  & \colhead{(Mag)}   
}
\startdata
  $2,457,106.48 \pm 0.09$  &  $0.11 \pm 0.04$  &
  $1.26 \pm 0.10$  & $13.36 \pm 0.10$ & $-19.42 \pm 0.11$  &  $31.89 \pm 0.04$   &
 $17.5 \pm 0.6^{\rm{a}}$   &  $0.035 \pm 0.033$  \\   
\enddata
\tablecomments{$^{\rm{a}}$ The error indicates a possible range in
$t_{\rm{rise}}$, which reflects the large uncertainty in  
the first light time as discussed in Section 3.2.}
\end{deluxetable}

\begin{deluxetable}{cccc}
\tablecaption{$R$-band Light Curve}
\tablewidth{0pt}
\tablehead{
\colhead{UT Date} & \colhead{Phase$^{\rm{a}}$} & \colhead{$F_{\nu}$} & \colhead{Error} \\
\colhead{2015}  &   \colhead{(Days)}   &  \colhead{($\mu$Jy)} & \colhead{($\mu$Jy)} 
}
\startdata
Dec 14.6146 &   -83.8480 &      -1.27 &       2.22\\
Dec 15.6540 &   -82.8086 &      -4.20 &       5.07\\
Dec 16.6511 &   -81.8115 &      -6.90 &       2.30\\
Dec 18.6542 &   -79.8084 &       6.09 &       3.03\\
Dec 19.7066 &   -78.7560 &      -2.78 &       5.92\\
Dec 20.6428 &   -77.8198 &       7.28 &       3.06\\
Dec 21.6767 &   -76.7859 &       0.61 &       6.17\\
Dec 31.5655 &   -66.8971 &      -0.62 &       4.95\\
Jan 07.6148 &   -59.8478 &      -0.29 &       4.27\\
Jan 08.5116 &   -58.9510 &       1.70 &       3.57\\
Jan 15.7366 &   -51.7260 &      -8.58 &       5.27\\
Jan 16.6335 &   -50.8291 &      -3.02 &       2.71\\
Jan 17.6600 &   -49.8026 &       4.55 &       4.29\\
Jan 18.6244 &   -48.8382 &       4.05 &       4.02\\
Jan 23.5471 &   -43.9155 &      -1.46 &       2.09\\
Jan 29.5524 &   -37.9102 &      -0.04 &       2.81\\
Jan 30.5336 &   -36.9290 &       6.96 &       3.65\\
Jan 31.6084 &   -35.8542 &      -4.02 &       4.97\\
Feb 01.4900 &   -34.9726 &     -12.15 &       4.52\\
Feb 04.5529 &   -31.9097 &      -7.81 &       6.94\\
Feb 07.5780 &   -28.8846 &      -5.60 &       3.97\\
Feb 08.5681 &   -27.8945 &       4.53 &       9.35\\
Feb 09.5751 &   -26.8875 &     -14.72 &       6.88\\
Feb 10.5929 &   -25.8697 &       0.53 &       5.45\\
Feb 11.5777 &   -24.8849 &      -1.36 &       6.46\\
Feb 13.6005 &   -22.8621 &       0.97 &       4.97\\
Feb 15.5596 &   -20.9030 &       0.98 &       3.61\\
Feb 16.5984 &   -19.8642 &      10.97 &       9.33\\
Feb 18.5182 &   -17.9444 &       2.74 &       3.18\\
Feb 19.5996 &   -16.8630 &      -2.24 &       5.31\\
Feb 21.4670 &   -14.9956 &       1.23 &       4.11\\
Feb 22.5007 &   -13.9619 &       4.88 &       2.91\\
Feb 23.5556 &   -12.9070 &       2.37 &       2.11\\
Feb 26.4913 &    -9.9713 &      -2.24 &       3.02\\
Mar 02.4899 &    -5.9727 &       0.75 &       6.69\\
Mar 03.5667 &    -4.8959 &       0.39 &      22.67\\
Mar 04.5186 &    -3.9440 &      -4.98 &       5.09\\
Mar 05.5475 &    -2.9151 &      15.73 &       7.65\\
Mar 06.5386 &    -1.9240 &      11.91 &       4.71\\
Mar 07.5108 &    -0.9518 &       4.77 &       5.74\\
Mar 08.4626 &     0.0000 &     114.08 &      12.76\\
Mar 08.4649 &     0.0023 &     126.01 &      13.00\\
Mar 08.4672 &     0.0046 &     119.56 &      12.33\\
Mar 09.4553 &     0.9927 &     358.43 &      14.01\\
Mar 09.4577 &     0.9951 &     356.78 &      15.11\\
Mar 09.4600 &     0.9974 &     378.09 &      15.19\\
Mar 09.5398 &     1.0772 &     438.13 &      23.31\\
Mar 09.5421 &     1.0795 &     392.65 &      20.46\\
Mar 09.5445 &     1.0819 &     389.40 &      23.68\\
Mar 09.7901$^{\rm{b}}$ &     1.3275 &     448.75 &      30.69\\
Mar 11.4424 &     2.9798 &    1341.53 &      72.83\\
Mar 11.4449 &     2.9823 &    1379.11 &      59.89\\
Mar 11.4474 &     2.9848 &    1424.30 &      60.31\\
Mar 11.4963 &     3.0337 &    1441.45 &      37.56\\
Mar 11.4988 &     3.0362 &    1404.75 &      35.08\\
Mar 11.5014 &     3.0388 &    1421.67 &      35.50\\
Mar 11.8010$^{\rm{b}}$ &     3.3384 &    1519.15 &      61.03\\
Mar 14.7630$^{\rm{b}}$ &     6.3004 &    4733.69 &     215.86\\
Mar 15.7460$^{\rm{b}}$ &     7.2834 &    6338.69 &     227.11\\
Mar 16.4303 &     7.9677 &    7613.77 &     181.86\\
Mar 16.4327 &     7.9701 &    7578.79 &     181.03\\
Mar 16.4352 &     7.9726 &    7516.23 &     179.53\\
Mar 16.5189 &     8.0563 &    7741.05 &     184.90\\
Mar 16.5213 &     8.0587 &    7762.47 &     185.42\\
Mar 16.5237 &     8.0611 &    7812.68 &     186.62\\
\enddata
\tablecomments{$^{\rm{a}}$ Days from the first detection of radioactively powered emission on 2015 March 8.46 (JD 2,457,089.9626).\\
$^{\rm{b}}$ Data taken in South Africa (Monard). All the other data
 come from LSGT observation.}
\end{deluxetable}

\clearpage

\clearpage

\begin{figure}[p]
\center{
\includegraphics[width=17.cm]{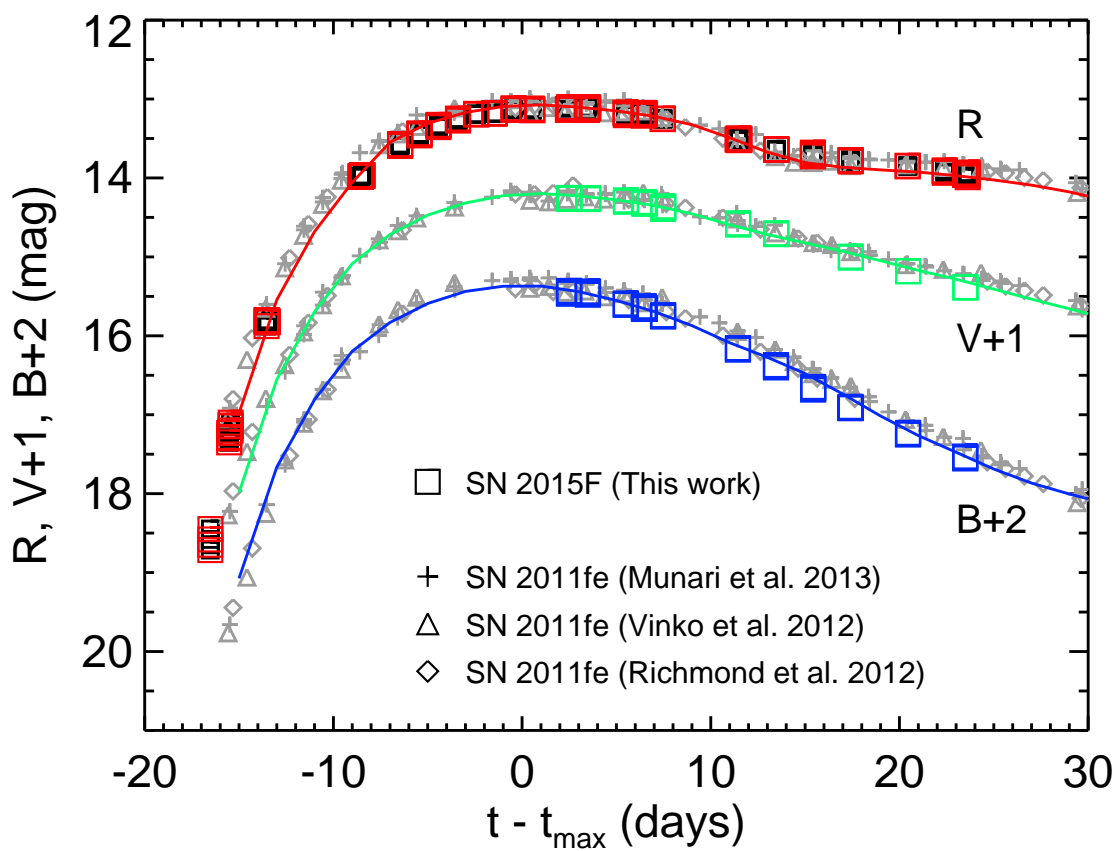}}
\caption{ 
 The long-term light curve of SN 2015F up to 22 days after the $B$-band maximum, along with the best-fit models and the SN 2011fe data points. The SN 2015F light curve suggests that it is
 a typical SN Ia, very similar in properties to SN 2011fe.}
\end{figure}

\begin{figure}[p]
\center{
\includegraphics[width=17.cm]{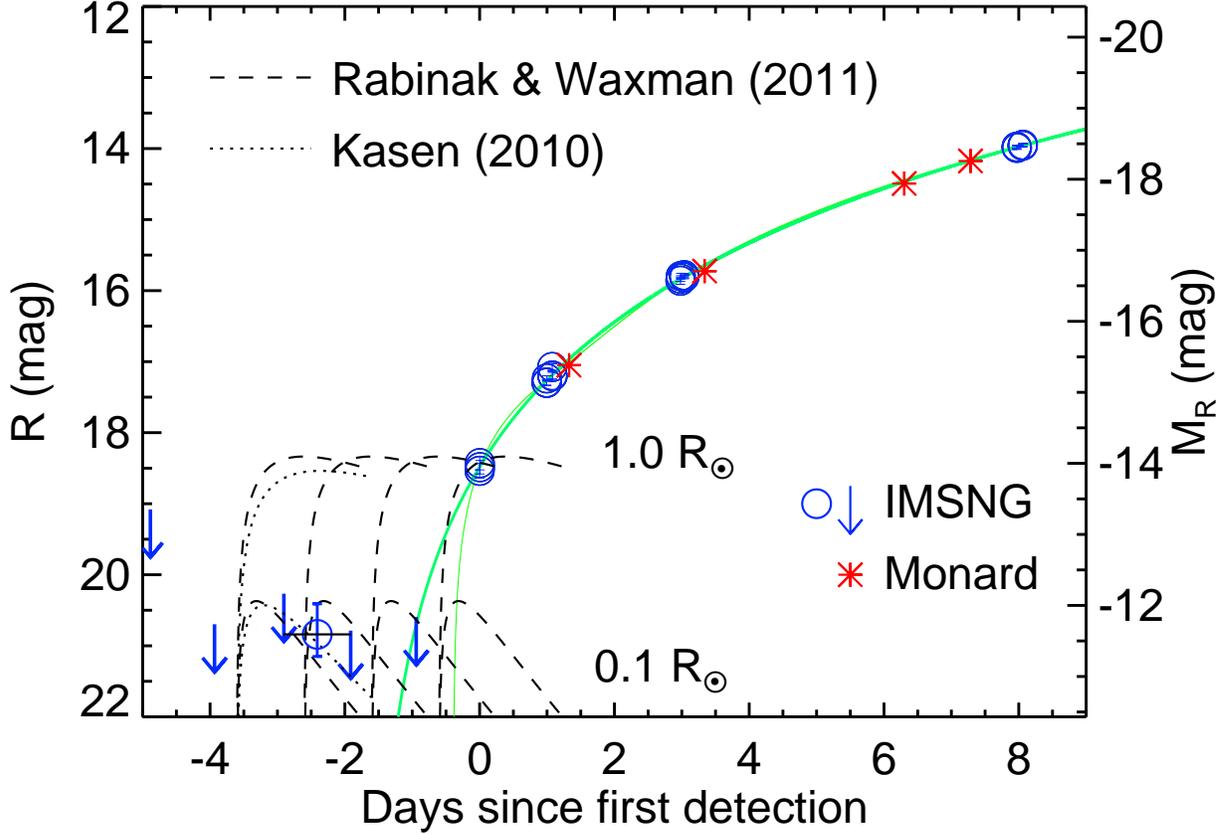}}
\caption{The $R$-band light curve of SN 2015F at the epochs 
between $-5$ days and $+8$ days from the first detection. 
 The blue circles and arrows (3$-\sigma$ upper limits)
 are for the LSGT observations and the red asterisks are for the data taken at South Africa.
 The average of the possible detections of shock-heated cooling emission is plotted at $-3$ days 
 (see Figure 4 and text for more detail).  
  The green solid lines indicate the best-fit results from 
 a single power-law (thick) and a broken power-law (thin) fits.  
 The dashed and the dotted lines are the predictions of two models (RW11 and K10)  
 of shock-heated cooling light curves for two progenitor star radii 
 (for the K10 model, the size of a companion star) of 0.1 and 1 $R_{\odot}$. 
  The model curves are plotted for four different explosion times at $-0.5$, $-1.5$, $-2.5$, and $-3.5$ days.
 The K10 model is for a viewing angle that is favorable to detect the early emission.
 }
\end{figure}

\begin{figure}[p]
\center{
\includegraphics[width=16.cm]{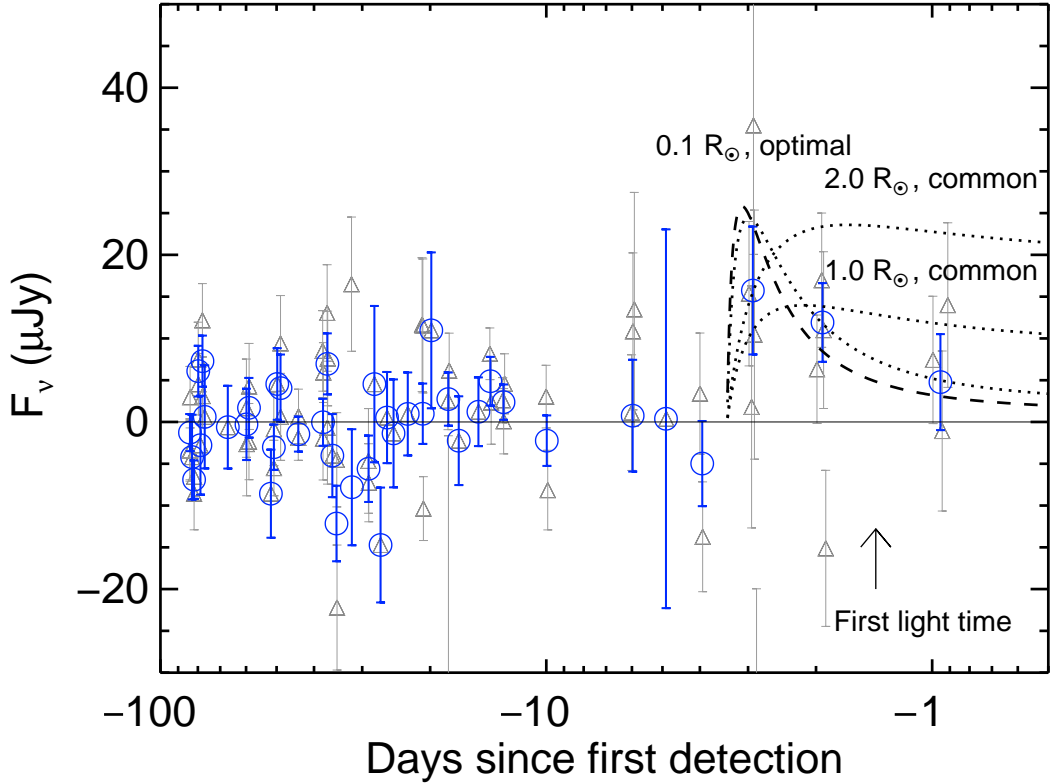}
\includegraphics[width=13.5cm]{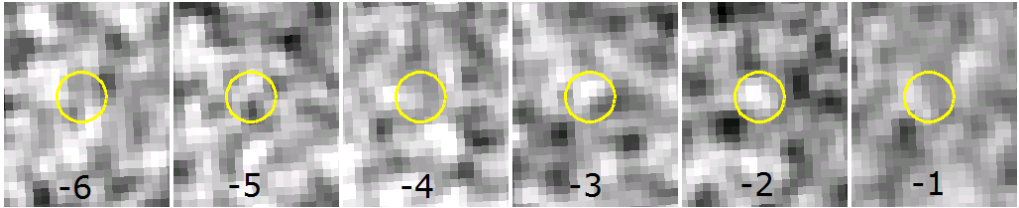}}
\caption{Possible detection of the early emission from shock-heated materials of SN 2015F. 
Top: The light curve from an aperture photometry ($3\farcs0$ diameter) 
that is performed at the SN 2015F location before and around the first light time. 
 The blue circles are the results from a stacked image of all images taken at each night, 
 and gray triangles are for measurements 
made on images taken at several different epochs each night (cadence of $\sim1.5$ hr). 
 Two notable events are recorded over two consecutive nights at $-3$ and $-2$ days, 
each with a 2-$\sigma$ significance. The RW11 (dashed line) and the K10 (dotted line) model  
predictions are shown for several different progenitor sizes. 
 For the K10 model, we plot the predictions for $R_{*} = 0.1~R_{\odot}$ under an optimal viewing angle, 
   and $R_{*} =$ 1 and 2 $R_{\odot}$ under a common viewing angle.  
  The signals at $-3$ and $-2$ days, if real, can be explained with a SD system having
  a companion star with $R_{*} \simeq 0.1$ -- $1.0 ~R_{\odot}$ or a prompt explosion of 
  a DD system ($R_{*} \sim 0.1~R_{\odot}$). This result excludes a 
  companion star with $R_{*} \gg 1~ R_{\odot}$ such as that of a subgiant or a red giant star, or
  a very small progenitor with $R_{*} \lesssim 0.01~R_{\odot}$ such as in a delayed detonation of a DD system.
 The first light time estimate from a single power-law fit is shown at $-1.6$ days. 
 Bottom: The difference images at the location of the SN 2015F which is marked with a $3\farcs0$ diameter 
yellow circle. The numbers in each panel shows the number of days before the first detection.
 The images are convolved with a $\sigma = 1$ pixel Gaussian kernel. }
\end{figure}

\begin{figure}[p]
\center{
\includegraphics[width=15.5cm]{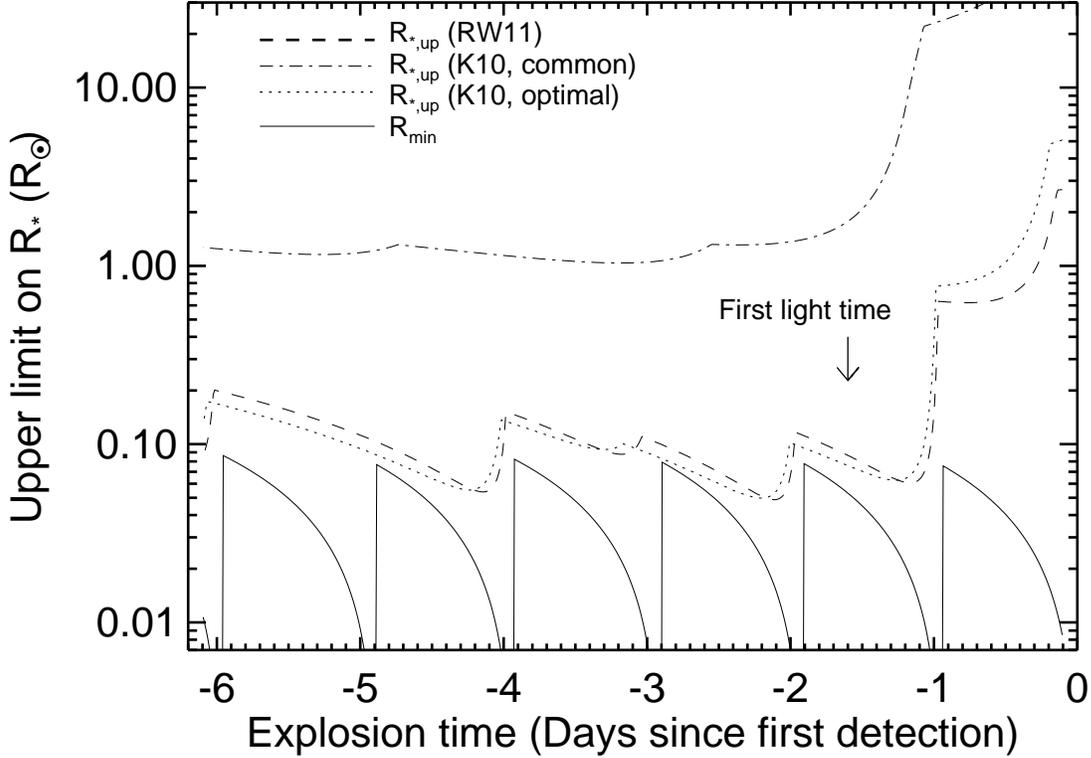}}
\caption{The upper limits on the progenitor radius as a function of the assumed
explosion time. The dashed line indicates the upper limits from the shock-heated 
cooling light curve of RW11, and the the solid line represents the minimal
radius we can explore with our data cadence. We also plot the upper limits from
the K10 model for an optimal 
(dotted line) and a more common viewing angle (dot-dashed line). 
The first light time from a single power-law fitting
is indicated at $-1.6$ days. These limits exclude a very large progenitor 
system ($R_{*} \gg 1~R_{\odot}$) unless the first light time is located within
one day from the first detection time.} 
\end{figure}

\end{document}